\documentclass{ws-procs9x6}
\input epsf.tex

\begin{document}


\title{EXPLORING THE 4d SUPERCONFORMAL ZOO}

\author{KEN INTRILIGATOR and BRIAN WECHT}
\address{UCSD PHYSICS DEPT, 9500 GILMAN DRIVE, LA JOLLA CA 92093}  


\maketitle

\abstracts{We discuss a new constraint for determining the superconformal $U(1)_R$ symmetry
of 4d $N=1$ SCFTs:
It is the unique one which locally maximizes $a(R)\equiv 3Tr R^3-Tr R$.  This
constraint comes close to proving the conjectured ``a-theorem" for $N=1$
SCFTs.  Using this ``a-maximization", exact results can now be obtained for previously inaccessible 4d
$N=1$ SCFTs.  We apply this method to a rich class of examples: 4d $N=1$
SQCD with added matter chiral superfields in the adjoint representation.  We classify a zoo
of SCFTs, finding that Arnold's ADE singularity classification arises in classifying
these theories via all possible relevant Landau-Ginzburg superpotentials.  We verify that
all RG flows  are indeed compatible with the ``a-theorem" conjecture, $a_{IR}<a_{UV}$, in
every case.}


There is an intuitive picture of renormalization group (RG) flows from the UV to the
IR as being like the flows of water down mountains, through valleys, and eventually
ending up in lakes.  The water height corresponds to the number of
``degrees of freedom'' of the theory, which is decreasing because of the coarse graining
removal of UV degrees of freedom in RG flow to the IR.   
Zamolodchikov \cite{Zamo} made this intuition precise in 2d by defining a  
$c$-function which counts the number of degrees of freedom of the theory, and proving that it monotonically decreases along RG flows to the IR.  In the far IR, the RG can flow to a
fixed point, which is a conformal field theory (CFT).
The c-function is stationary at the CFT endpoint of the flow, and in the 2d case
it becomes the central charge $c$ of the Virasoro algebra.  

The intuition about course graining suggests that such a 
monotonically decreasing c-function should exist for RG flows in any space-time dimension,
but a definitive proof above 2d has been elusive.  Cardy\cite{Cardy} has suggested that an  appropriate central charge for 4d CFTs (i.e. only the endpoints of the RG flow) is the 
coefficient, conventionally called $``a"$,
of the conformal anomaly $\langle T^\mu _\mu\rangle$ of the theory on a curved space-time background\footnote{On a general background $\langle T^\mu _\mu \rangle \sim a({\rm Euler})+
c({\rm Weyl ^2})$.  It is known from counterexamples that the coefficient $``c"$ of
the Weyl curvature squared term does 
{\it not} obey a 4d analog of the $c$-theorem: there are RG flows with $c_{IR}>c_{UV}$
\cite{Anselmi}.}
such as $S^4$.  This proposal is the conjectured
4d ``a-theorem'', that all 4d RG flows have $a_{IR}<a_{UV}$.  All known 4d RG flows
are compatible with this conjecture\footnote{Another conjectured c-function candidate is the coefficient of $T^4$ in the free-energy
per unit volume of the theory at finite temperature\cite{Appel}.  This  has some
advantages over the conformal anomaly: it is not restricted to even space-time dimensions, 
and it places tighter bounds on the maximum number of massless flavors where chiral symmetry breaking can occur.  But it has the disadvantage that it can only be readily computed in free
field theories.  In particular, supersymmetry does not help to compute it, so we will not discuss
this thermal c-function candidate further here.}.  We'll 
discuss some new RG flows and fixed points
which provide additional striking, non-trivial evidence for its validity.

4d asymptotically free gauge theories have $g(\mu)\rightarrow 0$ in the extreme UV, i.e.
free quarks and gluons, with $g$ increasing in the flow to the IR.  
Eventually, non-perturbative phenomena
can take over, e.g. confinement and a mass gap, and the theory in the far IR can be
a free field theory of composites, e.g. pions, or have no massless degrees of freedom
at all.  Another possibility is a non-trivial IR fixed point, an  
interacting CFT (no mass gap).  This happens if the RG flow drives the coupling
$g$ to a critical value $g_*$, where the beta functions vanish, $\beta (g_*)=0$.  
A scenario for realizing this phase was argued for
long ago\cite{GW,BZ}: if the theory is just barely asymptotically free, $\beta (g)$ can
have a non-trivial zero, $\beta (g_*)=0$, for $g_*\neq 0$   which is sufficiently weak
to justify the perturbative analysis.

Interacting 4d CFTs were once thought to be very
rare and exotic, perhaps only occurring in contrived examples.  However, we now know
of many such interacting RG fixed points. In this talk, we'll mention some
new ones, which we've recently found and explored.  

For the case of supersymmetric gauge theories, some exact results can be obtained, 
giving better insight into their non-perturbative dynamics.  These studies have dovetailed with the use of other
non-perturbative methods, which do not rely on supersymmetry, to
explore more general strongly interacting gauge theories.  A lesson learned from these
studies is that the non-Abelian Coulomb phase exists somewhat generically, provided 
that there are sufficiently many massless matter fields (but not too many so as to spoil asymptotic freedom). 

For example, in $N=1$ supersymmetric QCD, Seiberg\cite{Seiberg} argued that the theory flows to an interacting 4d $N=1$ SCFT for $N_f$ in the range 
${3\over 2}N_c<N_f<3N_c$.  More generally, 
 we suspect that $N=1$ SCFTs are fairly generic for matter content in the
range $T(G)<T(Matter)<3T(G)$ (there is also the more exotic possibility of the
free-magnetic phase, as seen in SQCD\cite{Seiberg}; see e.g. \cite{intseib} for a review).  
Here $T(Matter)$ and $T(G)$ are
the quadratic Casimirs of the matter and adjoint representations of the gauge group $G$,
with $T(Matter)<3T(G)$ for asymptotic freedom and $T(Matter)>T(G)$ to avoid 
 dynamically generated superpotentials, which would spoil conformal invariance.

Supersymmetry relates the stress tensor $T_{\mu \nu}$ to
a $U(1)_R$ current $R_\mu$: they are both in a current supermultiplet $T_{\alpha \dot \alpha}$ whose first component is the $U(1)_R$ current ($\alpha$ and $\dot \alpha$ are spinor indices). Let's call the charge associated with this $U(1)_R$ current $R_T$.  Away from RG fixed points, $R_T$ isn't really conserved: its non-conservation
is related by supersymmetry to non-zero $T_\mu ^\mu$ and lack of scale invariance.
As the theory RG flows to a SCFT, $R_T\rightarrow R_*$ which is conserved.  4d
SCFTs necessarily have a conserved $U(1)_{R_*}$ symmetry, with $U(1)_{R_*}\subset 
SU(2,2|1)$, the 4d superconformal group.  We refer to this $U(1)_{R_*}$
as the superconformal R-symmetry.  The supersymmetry relation 
$R_\mu \leftrightarrow T_{\mu \nu}$ implies many powerful results, for example:

$\bullet$ $\Delta({\mathcal O})\geq {3\over 2}|R_*({\mathcal O})|$ for all gauge invariant
spin zero operators ${\mathcal O}$, with $\Delta ({\mathcal O})$ the exact operator dimension and $R_*({\mathcal O})$ the operator's $U(1)_{R_*}$ charge.  
Chiral primary operators have $\Delta ({\mathcal O})={3\over 2}R_*({\mathcal O})$,
and additivity of the $U(1)_R$ charge for composite operators implies that they form
a closed OPE ring, with additive operator dimensions.   A unitarity bound, 
$\Delta ({\mathcal O})\geq 1$ for spin zero operators, implies that 
$R_*({\mathcal O}) \geq {2\over 3}$ for spin zero chiral primary operators, with 
$R_*({\mathcal O}) ={2\over 3}$ if and only if  ${\mathcal O}$ is a decoupled free field. 

$\bullet$   $R_*$ is anomaly free precisely if the 
exact NSVZ\cite{NSVZ} beta function vanishes, $\beta _{NSVZ}(g_*)=0$,
as is appropriate for a CFT. 

$\bullet$ The conformal anomalies $a$ and $c$ can be exactly related\cite{Anselmi} to the
superconformal R-symmetry's 
't Hooft anomalies:
$$a={3\over 32}(3{\mathrm Tr} R_*^3-{\mathrm Tr} R_*) \qquad c={1\over 32}(9{\mathrm Tr} R_*^3-5{\mathrm Tr} R_*).$$  This is extremely powerful, 
because 't Hooft anomaly matching \cite{thooft}
implies that such 't Hooft anomalies are 
constant along the RG flows\footnote{Here
we are supposing the the $R_*$ symmetry is conserved along the RG flow, rather than being
an accidental symmetry of the IR fixed point.  Away from the IR fixed point, $R_*$ differs
{}from the R-current in the $T_{\alpha \dot \alpha}$ supermultiplet. } and can thus be evaluated in the weakly coupled UV limit.  In what follows, we will rescale $a$ to eliminate
the factor of $3/32$.  

As an example, consider $SU(N_c)$ SQCD, with $N_f$ flavors in the SCFT range\cite{Seiberg} ${3\over 2}N_c<N_f<
3N_c$.  This theory has a unique
anomaly free $U(1)_R$ symmetry which commutes with all of the other flavor symmetries: 
the R-charges of the fundamentals are $R(Q_i)=R(\widetilde Q_i)=
1-(N_c/N_f)$.  This $U(1)_R$ is conserved along the entire RG flow, from
$g=0$ in the UV to the $g_*$ in the IR where $\beta (g_*)=0$, so it's natural to 
suppose that the superconformal $U(1)_{R_*}$ of the IR SCFT is this unique candidate (rather than being an accidental symmetry).   This yields the exact operator dimensions of chiral primary operators via $\Delta = {3\over 2}R$, 
 e.g. the gauge invariant mesons $M_{i\tilde j}=Q_i\widetilde Q_{\tilde j}$
have $\Delta (M)={3\over 2}R(M)=3(1-(N_c/N_f))$.   And, using 't Hooft 
anomaly matching, we   
exactly compute $a$ and $c$, even at strongly interacting RG fixed points, using the 
spectrum of the UV free field theory. 

More generally, to make use of the above powerful supersymmetry relations, one must
be able to identify precisely {\it which} possible R-symmetry is the special one, $R_*$,
of the superconformal algebra.   It is generally not uniquely fixed by the symmetries, as
it was in the SQCD example.  Indeed,
if the theory has a large group ${\mathcal F}$ of non-R global flavor symmetries, with charges $F_I$,
we can make a general R-symmetry by combining any initial R-symmetry, $R_0$,
with any linear combination of the flavor symmetries:
$$R_t = R_0 + \sum_I s_I F_I,$$
with the $s_I$ real parameters.  
The question, then, is to determine the particular 
values of the $s_I$ which produce the special $U(1)_{R_*}\subset SU(2,2|1)$.

We recently found a simple solution for this problem\cite{IW}: the superconformal
R-symmetry is the unique choice of the trial R-symmetry, $R_t$ as given above, which
locally maximizes 
$$a_{\mathrm trial}(s_I) = 3{\mathrm Tr} R_t^3 - {\mathrm Tr} R_t.$$
The value of $a_{trial}$ at this local maximum is then the central charge $a$ of the
SCFT, so we refer to this procedure for finding the R-charge as ``a-maximization."
(Because $a_{trial}$ is a cubic function, there is a unique local maximum, but no
global maximum.)  

We proved the above by showing  that the $T_{\mu \nu}\leftrightarrow R_\mu$ 
supersymmetry relation implies that 
$9 {\mathrm Tr} R_*^2 F_I = {\mathrm Tr} F_I$, from which it follows that the
superconformal $U(1)_R$ extremizes $a_{trial}$,  and also 
${\mathrm Tr}R_*F_I F_J < 0$, which implies that the extremum is
a local maximum.  The relation $9 {\mathrm Tr} R_*^2 F_I = {\mathrm Tr} F_I$
is obtained by relating the $F_I R_*^2$ three-current triangle anomaly to
another triangle anomaly, with the $R_*$ currents replaced by stress tensors.  
The relation ${\mathrm Tr}R_*F_I F_J < 0$ comes from relating the $F_IF_J R_*$
three-current triangle anomaly to the $F_I$, $F_J$, stress tensor three-point 
function, which is then related to the $F_I$, $F_J$ current-current two point
function by a general relation for 4d CFTs \cite{Osborn}.  
A unitarity condition, fixing the
sign of the current-current two-point function, then 
implies ${\mathrm Tr}R_*F_I F_J < 0$.

As a simple example to illustrate the a-maximization procedure, consider 
a free chiral superfield $\Phi$, with trial R-charge $r$.  We
compute  $a_{\mathrm trial} = 3(r-1)^3 - (r-1)$, which has a unique local
maximum at $r=2/3$, the correct value for a free field:
$\Delta = 3R/2 = 1$.   a-maximization for interacting
theories is essentially just as simple as this free-field example, the only
difference being that some constraints associated with the interactions
must be imposed.  For example, gauge interactions simply have the
effect of imposing the condition that the 
R-symmetry must be anomaly free.  

a-maximization also suggests a simple proof of the conjectured a-theorem, at least in
the supersymmetric context\cite{IW}.   Consider a general RG flow, from some CFT in the UV
to another CFT in the IR.  Often the flavor symmetry group of the IR
theory is a subgroup of that of the UV theory, ${\mathcal F}_{IR} \subset {\mathcal F}_{UV}$,
because the relevant deformations of the UV CFT broke some of the flavor symmetries.  
It then follows from a-maximization that  $a_{IR}<a_{UV}$, simply because
maximizing over a subset leads to a smaller value.  This proof is not fully general,
because there are examples where ${\mathcal F}_{IR}$ is not a subset of ${\mathcal F}_{UV}$,  because of accidental symmetries.   In all such examples, 
$a_{IR}<a_{UV}$ is still satisfied, but this loophole in the proof needs filling.  
Another concern is that perhaps maximizing over a subspace need not lead to a
smaller value, because it's only a local maximum.   This latter caveat can 
be dispensed with by a recent
work by Kutasov\cite{Kutasov:2003ux}, where the constraints associated with 
 ${\mathcal F}_{IR} \subset {\mathcal F}_{UV}$ are imposed with Lagrange multipliers
$\lambda$ and it's shown that $a(\lambda)$ is monotonically decreasing between
the UV and IR endpoints.  

We also note that  a-maximization always yields the R-charges as solutions
of quadratic equations with rational coefficients.  Thus, for any 4d ${\mathcal N} =1$
SCFT,  all of the operator R-charges, chiral primary operator dimensions,
and central charges $a$ and $c$ are always algebraic
numbers.  Specifically, they're solutions of quadratic equations with rational coefficients
(``quadratic irrational numbers").  In particular, these quantities can't depend on 
any continuous moduli.  

Finally, a general caution:  we must maximize 
$a_{trial}=3{\mathrm Tr} R_t^3-{\mathrm Tr} R_t$ over the complete space
of all possible trial R-symmetries, including all accidental symmetries.    
One situation where accidental
symmetries are readily apparent, and required, is when a gauge invariant chiral primary
operator, e.g. $M=\widetilde Q Q$, hits or appears to violate the unitarity bound, 
$R(M)\geq 2/3$.  $M$ then becomes a free field, with an 
accidental symmetry, $J_M$, under which only $M$ is charged.  The trial $R_t$ must then
include mixing with $J_M$: $R_{t,new}= R_{t,old}+s_MJ_M$, with the new parameter
$s_M$ again fixed by maximizing $a_{trial}$.  This has an important effect, leading to an additive correction to the quantity $a_{trial}$ to be maximized w.r.t. the other parameters $s_I$,  as was first discussed by Kutasov, Parnachev, and Sahakyan (KPS) \cite{KPS}.

a-maximization can be used to study previously mysterious theories.  As an example, consider $SU(N_c)$ SQCD with $0<N_f<2N_c$ fundamental flavors, along with an extra matter chiral superfield, $X$, in the adjoint representation.  This theory, with $W_{tree}=0$,
is believed to flow in the IR to an interacting ${\mathcal N}=1$ SCFT.  This was argued
for in \cite{intseib} by turning on $W_{tree}=\lambda Q X \widetilde{Q}$, finding massless
monopoles and dyons on submanifolds of a Coulomb branch, and noting that these
non mututally-local fields all become massless at the origin upon taking $\lambda \rightarrow 0$.    We refer to these IR SCFTs as
the $\widehat A (N_c,N_f)$ theories, or simply $\widehat A$.   We can also consider the theory with added superpotential  $W_{tree}={\mathrm Tr}X^{k+1}$ which, for particular
ranges of $N_f$ (depending on $k$), can drive a RG flow to different SCFTs in the IR, which 
we'll refer to as the $A_k$
SCFTs.  Finding the $U(1)_R$ symmetry of the $\widehat A$ SCFTs requires a-maximization,
while that of the $A_k$ theory is determined by $W_{tree}$ and the symmetries.  

Let's outline the salient features of the $\widehat
A$ and $A_k$ theories, obtained by KPS\cite{KPS}.
Write the trial $U(1)_R$ as $R(Q_i)=R(\widetilde Q_i)\equiv y$, $R(X)=(1-y)/x$, with
$x\equiv N_c/N_f$.  To simplify things, take $N_c$ and $N_f$ large, holding $x>{1\over 2}$ fixed.     For $x\approx {1\over 2}$, 
the theory is just barely asymptotically free, and the RG fixed
point is of the ``Banks-Zaks" type \cite{GW,BZ}, at weak gauge coupling $g_*\ll 1$.
As we increase $x$, the RG fixed point moves to larger and larger values of the gauge
coupling $g_*$.  a-maximization can be used to determine the exact R-charges $y(x)$,
for all $x$.  As expected, $y(x)$ and $R(X)$ decrease with $x$, corresponding to the negative anomalous dimensions of gauge interactions.

As $x$ increases, generalized mesons of the form $M_j = \widetilde{Q}X^j Q$
successively hit, and then appear to violate, the unitarity bound $R({\mathcal O})\geq 2/3$.  
Each time this happens, there is an associated accidental flavor symmetry which
must be included in the trial $R_t$, leading to added contributions to the 
quantity $a_{trial}$ which is to be maximized. 
Because of this, it's best to let a computer solve the
required a-maximizations, numerically obtaining $y(x)$.    
The $x\rightarrow \infty$ limit can be treated analytically, with the result that 
$R(Q)\rightarrow  (\sqrt{3} -1)/3 \approx 0.244$, and $R(X) \rightarrow  0$ in the limit
\cite{KPS}.  

Suppose that we sit at the $\widehat A$ RG fixed point, and then perturb
the theory by adding a superpotential  $W_{A_k}=\lambda {\mathrm Tr}
X^{k+1}$.  This deformation is relevant if $\Delta (W_{tree})<3$, i.e.
if $R(X)<2/(k+1)$.  Using the a-maximization results, it is seen 
that $W_{A_k}$ can be relevant for any value of $k$, arbitrarily large, provided
that $x$ is sufficiently large, $x>x^{min}_{A_k}$, with $x^{min}_{A_k}$ determined
by solving $R(X)=(1-y(x^{min}_{A_k}))/x^{min}_{A_k}=2/(k+1)$.
For any $x>x^{min}_{A_k}$, $W_{A_k}$ is a relevant deformation of
the $\widehat A$ SCFT, driving a RG flow away from
the $\widehat A$ SCFT to a new
SCFT, which we call $A_k$.  The $A_k$ SCFTs exist for $x$ in the range
$x^{min}_{A_k}<x<x^{max}_{A_k}=k$, with $x^{max}_{A_k}=k$ the stability bound
\cite{Kutdual}.  This $x$ range is non-empty,  $x_{A_k}^{min}<x^{max}_{A_k}$,  for all $k$.   

The results of KPS \cite{KPS} also clarified the meaning of the duality found in\cite{Kutdual,Kutasov:1995np,Kutasov:1995ss}
for the $A_k$ theories, which relates the original $SU(N_c)$ theory to a similar one
with gauge group $SU(kN_f-N_c)$ and some additional gauge singlet mesons and
superpotential, generalizing the duality of\cite{Seiberg}.  In particular, 
there is a non-empty ``conformal window" range of $x$, $x^{min}_{A_k}<x<k-\widetilde x^{min}_{A_k}$ for which $W_{tree}=  {\mathrm Tr}
X^{k+1}$ is relevant in the electric theory, and its analog in the magnetic dual is also
relevant.  This is where the electric and magnetic dual descriptions are both useful.

There are many non-trivial checks of the a-theorem conjecture which can be made
here, and $a_{IR}<a_{UV}$ is indeed satisfied for all RG flows. E.g. 
we can flow from the UV to the IR as: free UV$\rightarrow \widehat A\rightarrow
A_k\rightarrow A_{k'}$, with $k^\prime < k$.  The a-theorem would then predict
$a_{free UV}(x)>a_{\widehat A}(x)>a_{A_k}(x)>a_{A_{k'}}(x)$.   Interestingly, $a_{A_k}(x)>a_{A_{k'}}(x)$ had some apparent violations for small $x$ \cite{Anselmi}. 
But using the new a-maximization results for $x^{min}_{A_k}$ it is now seen that
there never actually is a violation of $a_{IR}<a_{UV}$ in any of those examples.
The apparent violation of $a_{IR}<a_{UV}$ always occurred for $x$ 
 outside of the range $x>x^{min}_{A_k}$ needed for the $A_k$ SCFT to exist, so there's 
no a-theorem violating flow after all\cite{KPS}.

In our work \cite{IW2}, we generalized these examples by considering SQCD with $N_f$
fundamental flavors, $Q_i$ and $\widetilde Q_i$, along with $N_a=2$ adjoint matter flavors,
$X$ and $Y$.  As in \cite{intseib}, considering deformations suggests 
that this theory, with $W_{tree}=0$, flows to
an interacting ${\mathcal N}=1$ SCFT for all $N_f$  in the asymptotically free range $0\leq N_f<N_c$, which we write as $x\equiv N_c/N_f>1$.  
We'll call these SCFTs $\widehat O(N_c, N_f)$, and again consider $N_c$ and $N_f$
large, with $x$ fixed, to simplify things.  For $x\approx 1$, the $\widehat O$ IR fixed point
is at weak coupling, and as $x$ increases, the $\widehat O$ IR fixed point is at stronger
coupling.  We use a-maximization to solve for the R-charges for all $x$, taking the anomaly
free trial R-symmetry to be: $R_t(Q)=R_t(\widetilde Q)\equiv 
y$ and $R_t(X)=R_t(Y)={1\over 2}(1+{1-y\over x})$.
Plugging this into $a_{trial}=3{\mathrm Tr} R_t^3-
{\mathrm Tr}R_t$ and maximizing w.r.t. $y$ gives $y(x)=1+\left(3(8x^2-1)\right) ^{-1}\left(
3x-2x\sqrt{26x^2-1}\right)$.   We obtain a relatively simple closed form expression for
 $y(x)$ for the $\widehat O$ SCFTs because no operators hit the unitarity bound, 
so no additional contributions to
$a_{trial}$, associated with accidental symmetries are needed (though we can't
definitively rule them out).  We thus find that $R(Q)$ and $R(X)$ and $R(Y)$ are
decreasing functions of $x$, and for $x\rightarrow \infty$ we obtain $R(Q)\rightarrow
1-\sqrt{26}/12\approx 0.575$, and $R(X)=R(Y)\rightarrow {1\over 2}$.   

Using the above result, we can classify the relevant superpotential deformations
of the $\widehat O$ SCFT, which could drive RG flows from $\widehat O$ to various 
new SCFTs in the IR.  Let's consider superpotentials involving only the adjoints (there are additional ones using operators which
include the quarks).  Using our result for the
$\widehat O$ SCFT that  $R(X)=R(Y)>{1\over 2}$ for all $x$, we see
that $W={\mathrm Tr} X^kY^\ell$ is only a relevant deformation of the $\widehat O$
SCFT if $k+\ell\leq 3$.    The complete list of such relevant superpotential deformations 
(modulo field redefinitions)
of $\widehat O$ is thus $W_{\widehat A}={\mathrm Tr}Y^2$, 
$W_{\widehat D}={\mathrm Tr} XY^2$, and $W_{\widehat E}={\mathrm Tr} Y^3$.  Each 
of these drives $\widehat O$ to new SCFTs, which we name $\widehat A$, $\widehat D$, and $\widehat E$ respectively, for all $N_f$ in the range $0<N_f<N_c$.  
The $\widehat A$ case is that 
considered by \cite{KPS} and reviewed above.  

We can now use a-maximization to analyze the new $\widehat D$ and 
$\widehat E$ SCFTs.   E.g. for
the $\widehat D$ RG fixed points, we impose $R(XY^2)=2$, and the 
trial $U(1)_R$ charges are $R(Q)=R(\widetilde Q)\equiv y$, $R(X)=2(1-y)/x$, 
$R(Y)=1+(y-1)/x$.  Here we find that operators
do hit the unitarity bound, so their accidental symmetries must be accounted for 
in $a_{trial}$, in analogy with the $\widehat A$ case of \cite{KPS}.  Because of this,
we solved for $y(x)$ numerically. The asymptotic values in the limit 
$x\rightarrow \infty$ are $y\rightarrow -1/8$, $R(Y)\rightarrow 1$, and $R(X)\rightarrow 0$.  
 
Using these results for the $\widehat D$ SCFT, we can now classify its relevant
superpotential deformations.  Since 
$R(X)\rightarrow 0$ for $x\rightarrow \infty$, we find that $\Delta W={\mathrm Tr}X^{k+1}$
can be a relevant deformation of the $\widehat D$ SCFT, for arbitrarily large $k$,
provided that $x$ is sufficiently large, $x>x_{D_{k+2}}^{min}$.  We solved for
$x_{D_{k+2}}^{min}$ by using our results for the $\widehat D$ SCFT to determine
the $x$ where $R(X)=2/(k+1)$: we find  
$x^{min}_{D_5}=2.09$, $x^{min}_{D_6}=3.14$, etc., with $x^{min}_{D_{k+2}}
\rightarrow {9\over 8}k$ for $k\rightarrow \infty$.  When $x>x_{D_{k+2}}^{min}$, 
the relevant  $\Delta W$ drives a RG flow from the $\widehat D$ SCFTs to new SCFTs, which we name $D_{k+2}$.  
More generally, we can get to the $D_{k+2}$ SCFTs, provided that $x>x^{min}_{D_{k+2}}$,
by perturbing $\widehat O$ by  $W_{D_{k+2}}=\lambda _1 {\mathrm Tr} X^{k+1}+
\lambda _2 {\mathrm Tr}XY^2$.  (The 
$D_{k+2}$ SCFT also must satisfy $x<x^{max}_{D_{k+2}}=
3k$, to prevent  generating a $W_{dyn}$.)

Continuing this process of using a-maximization to analyze each SCFTs, classifying
their relevant superpotential deformations, and then using these deformations to
flow to new SCFTs,
we obtain a classification of the SCFTs that are obtainable from $\widehat O$ via
the following superpotential deformations (again, there are additional ones making use of operators involving the quarks): $W_{\widehat O}=0$, $W_{\widehat A}={\mathrm Tr}Y^2$, 
$W_{\widehat D}={\mathrm Tr} XY^2$, $W_{\widehat E}={\mathrm Tr} Y^3$, 
$W_{A_k}={\mathrm Tr}(X^{k+1}+Y^2)$, $W_{D_{k+2}}={\mathrm Tr}(X^{k+1}+XY^2)$,
$W_{E_6}={\mathrm Tr} (Y^3+X^4)$, $W_{E_7}={\mathrm Tr} (Y^3+YX^3)$, and 
$W_{E_8}={\mathrm Tr}(Y^3+X^5).$     Each of these LG superpotentials drives a RG
flow to a corresponding interacting SCFT in the IR, with the $A_k$, $D_{k+2}$,
$E_6$, $E_7$, and $E_8$ cases existing only if $x$ is sufficiently large.  

We note that the above superpotential
classification agrees with Arnold's singularity classification.  That classification also
appeared, among other contexts,  in the classification  of 2d SCFTs having $c<3$ \cite{Martinec,Vafa}.  The appearance of Arnold's ADE in our 4d case is a new guise,
which perhaps has a deeper connection to other occurrences of the ADE series in
string theory and mathematics.  This is a topic for future exploration. 

The map of
the possible flows between these SCFTs is:
\bigskip
\centerline{\epsfxsize=0.30\hsize\epsfbox{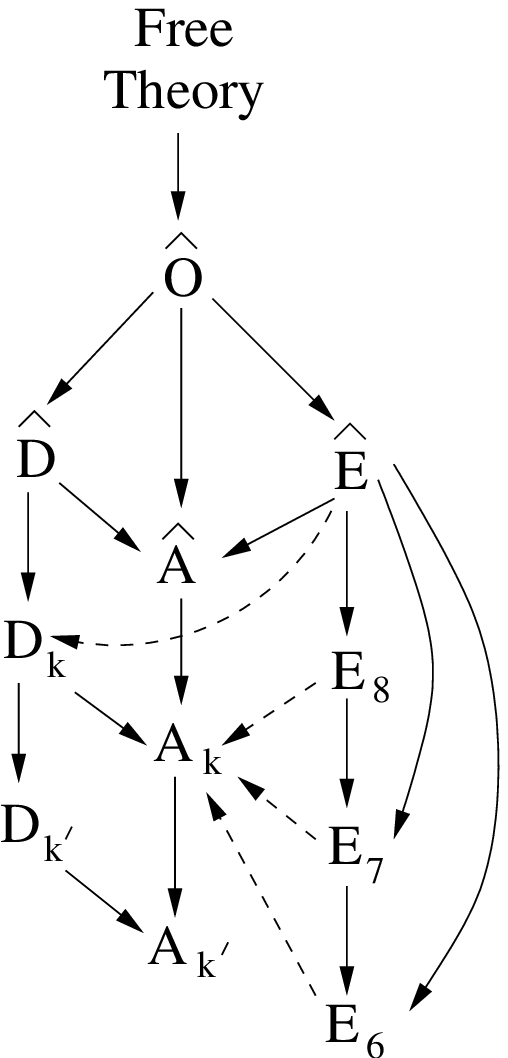}}
\centerline{\sl \baselineskip=8pt {\bf Figure 1:}
{\sl The map of possible flows between fixed points.}}
\centerline{\sl Dotted lines indicate flow to a particular
value
of $k$.}
\bigskip

Every arrow in the diagram corresponds to a RG flow, from a SCFT in the
UV to another SCFT in the IR.  For each of these flows, we used a-maximization
to exactly compute $a_{UV}$ and $a_{IR}$, and we verified that $a_{IR}<a_{UV}$
is indeed satisfied for every such RG flow.  E.g. $a_{E_6}(x)<a_{E_7}(x)<a_{E_8}(x)<a_{\widehat
E}(x)<a_{\widehat O}(x)<a_{free\ UV}(x)$.  There are some apparent violations for small $x$, e.g. 
it appears that $a_{E_7}(x)<a_{E_6}(x)$ for $x<3.16$, but there is actually no violation
of the a-theorem conjecture afterall, because we find that that $E_7$ RG fixed point
only exists if $x>x^{min}_{E_7}\approx 4.12$.  Likewise, all other such potential violations occur outside of the
range where the RG flow can exist.  

Our new results also give some insight into a proposed duality \cite{Brodie} for 
the theory with $W_{D_{k+2}}= {\mathrm Tr} (X^{k+1}+ XY^2)$, relating the original
``electric" $SU(N_c)$ theory to a ``magnetic" $SU(3kN_f-N_c)$ dual.  We find 
different phases depending on $x\equiv N_c/N_f$.  The duality is most
useful in the $D_{k+2}$ conformal window: $x^{min}_{D_{k+2}}<x<3k-
\widetilde x^{min}_{D_{k+2}}$, where $W_{D_{k+2}}$ is relevant in both the
electric and magnetic theories.  It's interesting that this range is non-empty
for all $k$, e.g. for large $k$ we find $x^{min}_{D_{k+2}}\approx {9\over 8}k$ and
$3k-\widetilde x^{min}_{D_{k+2}}\approx 1.8962k$.  Both dual descriptions are good
in the conformal window.  Outside of the conformal window, the theory can instead flow
to the electric or magnetic $\widehat D$ SCFT, and the other dual description is then
not useful.

The a-maximization methods used here can be applied to many other 4d supersymmetric gauge theories, and there are many other superconformal zoos remaining to be explored.



\section*{Acknowledgments}
KI thanks the organizers for the invitation to this fun conference.
This work was partially supported by
DOE-FG03-97ER40546.



\begin{thebibliography}{0}



\bibitem{Zamo}
A.~B.~Zamolodchikov,
JETP Lett.\  {\bf 43}, 730 (1986)
[Pisma Zh.\ Eksp.\ Teor.\ Fiz.\  {\bf 43}, 565 (1986)].

\bibitem{Cardy}
J.~L.~Cardy,
Phys.\ Lett.\ B {\bf 215}, 749 (1988).

\bibitem{Anselmi}
D.~Anselmi, D.~Z.~Freedman, M.~T.~Grisaru and A.~A.~Johansen,
Nucl.\ Phys.\ B {\bf 526}, 543 (1998)
[arXiv:hep-th/9708042].    

D.~Anselmi, J.~Erlich, D.~Z.~Freedman and A.~A.~Johansen,
Phys.\ Rev.\ D {\bf 57}, 7570 (1998)
[arXiv:hep-th/9711035].


\bibitem{Appel}
T.~Appelquist, A.~G.~Cohen and M.~Schmaltz,
Phys.\ Rev.\ D {\bf 60}, 045003 (1999)
[arXiv:hep-th/9901109].

\bibitem{GW}
D.~J.~Gross and F.~Wilczek,
Phys.\ Rev.\ D {\bf 9}, 980 (1974).

\bibitem{BZ}
T.~Banks and A.~Zaks,
Nucl.\ Phys.\ B {\bf 196}, 189 (1982).

\bibitem{Seiberg}
N.~Seiberg,
Nucl.\ Phys.\ B {\bf 435}, 129 (1995)
[arXiv:hep-th/9411149].

\bibitem{thooft}
G. ' t Hooft, {\it Recent Developments in Gauge Theories} 
Plenium Press {\bf 135} (1980), reprinted in
{\it Unity of Forces in the Universe Vol II} A. Zee ed.
World Scientific {\bf 1004} (1982)


\bibitem{intseib}
K.~A.~Intriligator and N.~Seiberg,
Nucl.\ Phys.\ Proc.\ Suppl.\  {\bf 45BC}, 1 (1996)
[arXiv:hep-th/9509066].

\bibitem{NSVZ}
V.~A.~Novikov, M.~A.~Shifman, A.~I.~Vainshtein and V.~I.~Zakharov,
Nucl.\ Phys.\ B {\bf 229}, 381 (1983).

\bibitem{Arnold}
V.I. Arnold, S.M. Gusein-Zade and
A.N. Varchenko, {\it Singularities of Differentiable Maps} volumes
1 and 2, Birkhauser (1985), and references therein.




\bibitem{IW}
K.~Intriligator and B.~Wecht,
Nucl.\ Phys.\ B {\bf 667}, 183 (2003)
[arXiv:hep-th/0304128].

\bibitem{Osborn}
H.~Osborn,
Annals Phys.\  {\bf 272}, 243 (1999)
[arXiv:hep-th/9808041].


\bibitem{KPS}
D.~Kutasov, A.~Parnachev and D.~A.~Sahakyan,
JHEP {\bf 0311}, 013 (2003)
[arXiv:hep-th/0308071].

\bibitem{Kutdual}
D.~Kutasov,
Phys.\ Lett.\ B {\bf 351}, 230 (1995)
[arXiv:hep-th/9503086].

\bibitem{Kutasov:1995np}
D.~Kutasov and A.~Schwimmer,
Phys.\ Lett.\ B {\bf 354}, 315 (1995)
[arXiv:hep-th/9505004].

\bibitem{Kutasov:1995ss}
D.~Kutasov, A.~Schwimmer and N.~Seiberg,
Nucl.\ Phys.\ B {\bf 459}, 455 (1996)
[arXiv:hep-th/9510222].

\bibitem{Brodie}
J.~H.~Brodie,
Nucl.\ Phys.\ B {\bf 478}, 123 (1996)
[arXiv:hep-th/9605232].

\bibitem{Intriligator:1994sm}
K.~A.~Intriligator and N.~Seiberg,
Nucl.\ Phys.\ B {\bf 431}, 551 (1994)
[arXiv:hep-th/9408155].

\bibitem{Intriligator:1995id}
K.~A.~Intriligator and N.~Seiberg,
Nucl.\ Phys.\ B {\bf 444}, 125 (1995)
[arXiv:hep-th/9503179].


\bibitem{IW2}
K.~Intriligator and B.~Wecht,
Nucl.\ Phys.\ B {\bf 677} (2004) 223
[arXiv:hep-th/0309201].

\bibitem{Kutasov:2003ux}
D.~Kutasov,
[arXiv:hep-th/0312098].


\bibitem{Martinec}
E.~J.~Martinec,
Phys.\ Lett.\ B {\bf 217}, 431 (1989).

\bibitem{Vafa}
C.~Vafa and N.~P.~Warner,
Phys.\ Lett.\ B {\bf 218}, 51 (1989).


\end{thebibliography}
\end{document}